\documentclass[12pt,a4paper]{article}
\usepackage{amsmath,amssymb}
\usepackage{graphicx}
\begin{document}
\title{
A Diagrammer's Note on 
Superconducting Fluctuation Transport 
for Beginners: \\
Supplement. Jonson-Mahan Transmutation 
}

\author{
O. Narikiyo
\footnote{
Department of Physics, 
Kyushu University, 
Fukuoka 812-8581, 
Japan}
}

\date{
(Sep. 17, 2013)
}

\maketitle
\begin{abstract}
The Ward identity for the heat current vertex is illustrated 
in terms of Feynman diagrams. 
The Jonson-Mahan transmutation, 
the way how the kinetic energy at the heat current vertex 
for the free propagator is transmuted 
into the frequency for the full propagator, 
is the key of the illustration. 
\end{abstract}
\vskip 20pt

\section{Introduction}

This Note is the Supplement 
to the series of Notes\footnote{The Notes are quoted as [I] and [II] 
where [I] $\equiv$ arXiv:1112.1513 and [II] $\equiv$ arXiv:1203.0127. } 
on the superconducting fluctuation transport 
where the quasi-particle transport is also discussed. 
The necessity of this Supplement is noticed in \S 14 of [I]. 

I think that the Jonson-Mahan transmutation \cite{JM}, 
the way how the kinetic energy at the heat current vertex 
for the free propagator is transmuted into the frequency 
for the full propagator, is the key to understand the Ward identity 
for the heat current vertex 
but is not recognized by most authors 
and that the ignorance of it 
leads to some confusions seen in the literatures.\footnote{
See the footnote for (\ref{wrong}) in this Supplement. 
For example, 
in the perturbational calculation in the Appendix-B 
of Michaeli and Finkel'stein: Phys. Rev. B {\bf 80}, 115111 (2009), 
the frequency is erroneously employed for the free vertex. 
Originally, the definition of the free vertex (2) in this work is incorrect. 
Including this work the use of $ c^\dag_{{\bf p}\sigma}(\varepsilon_n) $ 
is misleading. 
Since it is the Fourier transform of 
$e^{K\tau} c^\dag_{{\bf p}\sigma} e^{-K\tau}$, 
it is highly nonlinear in the operator set 
$\{ c^\dag_{{\bf p'}\sigma'}, c_{{\bf p''}\sigma''} \} $ 
in general. 
However, for example, $ c^\dag_{{\bf p}\sigma}(\varepsilon_n) $ in (18) 
of Kontani: Phys. Rev. B {\bf 67}, 014408 (2003) 
is erroneously interpreted as a simple one-body operator. 
Correctly, it is a complex many-body operator. 
The Fourier transform should be introduced for the propagators 
as usual and as the discussion on the Jonson-Mahan formula in [I]. 
As another example, 
the perturbational calculation of the heat current vertex for Cooper pairs, 
employing the frequency for the free vertex, 
by Ussishkin: Phys. Rev. B {\bf 68}, 024517 (2003) 
will be criticized in the next Supplement noticed in \S 15 in [I]. } 

In the following 
I illustrate in terms of Feynman diagrams 
the way how the Ward identity is satisfied for the heat current vertex. 
The Jonson-Mahan transmutation \cite{JM} is the key of the illustration. 

The symbols that have appeared in [I] and [II] 
are used here without explanations. 

\section{Heat-Current Operator}

The charge current of electrons, (59) in [I], 
is Fourier-transformed into 
\begin{equation}
{\bf j}^e({\bf k}) 
= e \sum_{\bf p} \sum_\sigma 
{ {\bf v}_{\bf p} + {\bf v}_{\bf p+k} \over 2 } 
c_{{\bf p}\sigma}^\dag c_{{\bf p+k}\sigma}, 
\label{j^e(k)} 
\end{equation}
which is equivalent to (73) in [I]. 
Here the operator is transformed as (75) in [I]. 
The uniform current is 
\begin{equation}
{\bf j}^e({\bf k} \!=\! 0) 
= e \sum_{\bf p} \sum_\sigma {\bf v}_{\bf p} 
c_{{\bf p}\sigma}^\dag c_{{\bf p}\sigma}. 
\label{j^e(0)} 
\end{equation}

The heat current of electrons, (65) in [I], 
is Fourier-transformed into 
\begin{equation}
{\bf j}^Q({\bf k}) 
= \sum_{\bf p} \sum_\sigma 
{     \xi_{\bf p} +     \xi_{\bf p+k} \over 2 } 
{ {\bf v}_{\bf p} + {\bf v}_{\bf p+k} \over 2 } 
c_{{\bf p}\sigma}^\dag c_{{\bf p+k}\sigma}, 
\label{j^Q(k)-free} 
\end{equation}
for free electrons.\footnote{
For free electrons 
\begin{equation}
{\bf v}_{\bf p} = {{\bf p} \over m}, 
\ \ \ \ \ \ \ \ \ \ \ \ 
\xi_{\bf p} = {{\bf p}^2 \over 2m} - \mu. 
\nonumber
\end{equation}
} 
The uniform current is 
\begin{equation}
{\bf j}^Q({\bf k} \!=\! 0) 
= \sum_{\bf p} \sum_\sigma \xi_{\bf p} {\bf v}_{\bf p} 
c_{{\bf p}\sigma}^\dag c_{{\bf p}\sigma}. 
\label{j^Q(0)-free} 
\end{equation}
Thus each component $c_{{\bf p}\sigma}^\dag c_{{\bf p}\sigma}$ 
carries the charge $e$ in (\ref{j^e(0)}) 
and the kinetic energy $\xi_{\bf p}$ in (\ref{j^Q(0)-free}). 

For interacting electrons 
the heat current also carries the interaction energy as follows. 
The Fourier transform of (88) in [I] is given as 
\begin{equation}
{\bf j}^Q({\bf k} \!=\! 0) = 
\lim_{\tau' \rightarrow \tau} {1 \over 2} 
\bigg( {\partial \over \partial \tau} - {\partial \over \partial \tau'} \bigg) 
\sum_{\bf p} {\bf v}_{\bf p} \Big( 
c_{{\bf p}\uparrow}^\dag(\tau) c_{{\bf p}\uparrow}(\tau') + 
c_{{\bf p}\downarrow}^\dag(\tau) c_{{\bf p}\downarrow}(\tau') \Big). 
\label{j^Q(tau)} 
\end{equation}
The time-dependence is determined by (8) in [I] with 
\begin{equation}
K = K_0 + V, 
\end{equation}
where 
\begin{equation}
K_0 = \sum_{\bf p} \xi_{\bf p} \Big( 
c_{{\bf p}\uparrow}^\dag c_{{\bf p}\uparrow} + 
c_{{\bf p}\downarrow}^\dag c_{{\bf p}\downarrow} \Big), 
\end{equation}
and 
\begin{equation}
V = \lambda \sum_{\bf p'} \sum_{\bf p''} \sum_{\bf q} 
c_{{\bf p'+q}/2\uparrow}^\dag  c_{{\bf -p'+q}/2\downarrow}^\dag 
c_{{\bf -p''+q}/2\downarrow}  c_{{\bf p''+q}/2\uparrow}. 
\end{equation}
These are equal to (2), (3) and (5) in [I] when we put $\lambda = -g$. 
The time-derivative is calculated as 
\begin{equation}
{\partial \over \partial \tau} c_{{\bf p}\uparrow}^\dag(\tau) 
= \big[ K, c_{{\bf p}\uparrow}^\dag \big] 
= \xi_{\bf p} c_{{\bf p}\uparrow}^\dag 
+ \lambda \sum_{\bf p'} \sum_{\bf p''} \sum_{\bf q} 
c_{{\bf p'+q}/2\uparrow}^\dag  c_{{\bf -p'+q}/2\downarrow}^\dag 
c_{{\bf -p''+q}/2\downarrow}  \delta_{{\bf p},{\bf p''+q}/2}, 
\end{equation}
\begin{equation}
{\partial \over \partial \tau'} c_{{\bf p}\uparrow}(\tau') 
= \big[ K, c_{{\bf p}\uparrow} \big] 
= - \xi_{\bf p} c_{{\bf p}\uparrow} 
- \lambda \sum_{\bf p'} \sum_{\bf p''} \sum_{\bf q} 
c_{{\bf -p'+q}/2\downarrow}^\dag  c_{{\bf -p''+q}/2\downarrow} 
c_{{\bf p''+q}/2\uparrow}  \delta_{{\bf p},{\bf p'+q}/2}, 
\end{equation}
and so on. 
Thus (\ref{j^Q(tau)}) leads to 
\begin{align}
{\bf j}^Q({\bf k} \!=\! 0) 
& = \sum_{\bf p} \sum_\sigma \xi_{\bf p} {\bf v}_{\bf p} 
c_{{\bf p}\sigma}^\dag c_{{\bf p}\sigma} 
\nonumber \\ 
& + {\lambda \over 2} \sum_{\bf p'} \sum_{\bf p''} \sum_{\bf q} 
\big( {\bf v}_{{\bf p''+q}/2}  + {\bf v}_{{\bf p'+q}/2} 
    + {\bf v}_{{\bf -p''+q}/2} + {\bf v}_{{\bf -p'+q}/2} \big) 
\nonumber \\ 
& \ \ \ \ \ \ \ \ \ \ \ \ \ \ \ \ \ \ \ \ \ \times
c_{{\bf p'+q}/2\uparrow}^\dag  c_{{\bf -p'+q}/2\downarrow}^\dag 
c_{{\bf -p''+q}/2\downarrow}  c_{{\bf p''+q}/2\uparrow}. 
\label{j^Q=K+V} 
\end{align}
This  quartic interaction term in the heat current, 
which does not exist in the quadratic charge current, 
makes the following discussions complicated. 

\vskip 12mm
\begin{figure}[htbp]
\begin{center}
\includegraphics[width=8.0cm]{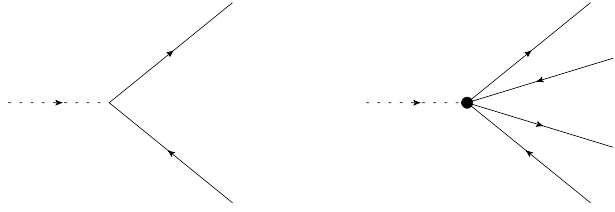}
\vskip 4mm
\caption{Diagrams for un-renormalized current vertex. 
In the case of heat current the right-hand side of (\ref{j^Q=K+V}) 
is represented by these two diagrams. 
In the case of charge current (\ref{j^e(0)}) 
is represented only by the left diagram. }
\label{fig:J}
\end{center}
\end{figure}

\section{Diagrammatics for Charge Vertex}

\begin{figure}[htbp]
\begin{center}
\includegraphics[width=9.0cm]{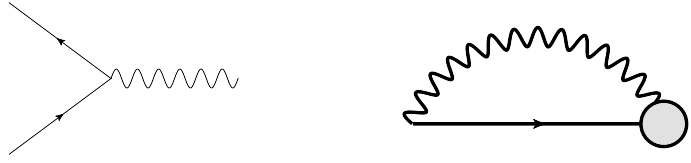}
\vskip 4mm
\caption{Diagrams for electron-boson interaction (left) and 
self-energy (right). 
The line with an arrow represents 
the free propagator $G_0$ (thin) or 
the full propagator $G$ (thick) for electrons. 
The wavy line represents 
the free propagator $D_0$ (thin) or 
the full propagator $D$ (thick) for bosons. 
The shaded circle is the full interaction vertex $\Lambda$. }
\label{fig:Sigma}
\end{center}
\end{figure}

Let us consider the Ward identity for the charge current vertex 
in terms of electron-boson interaction\footnote{
This interaction has the same structure as the electron-phonon interaction. 
See, for example, (4-26) and (4-27) in \cite{Sch}. 
Neglecting the polarization of the phonon 
the interaction is expressed as 
\begin{equation}
V_{\rm phonon} = \sum_{\bf p'} \sum_{\bf q} \sum_\sigma 
g_{\bf q} \varphi_{\bf q} 
c_{{\bf p'+q}\sigma}^\dag  c_{{\bf p'}\sigma}, 
\nonumber
\end{equation}
with 
\begin{equation}
\varphi_{\bf q} = a_{\bf q} + a_{\bf -q}^\dag, 
\nonumber
\end{equation}
which is the operator for the absorption/emission of the phonon. } 
\begin{equation}
V_{\rm boson} = \lambda \sum_{\bf p'} \sum_{\bf q} \sum_\sigma \chi_{\bf q} 
c_{{\bf p'+q}\sigma}^\dag  c_{{\bf p'}\sigma}, 
\end{equation}
shown in Fig.~\ref{fig:Sigma}-(left). 
The interaction between two electrons is represented 
by the exchange of the boson. 
The self-energy $\Sigma$ of electrons\footnote{
Using the faithful representation mentioned in the footnote 13 of [II] 
\begin{equation}
-\Sigma(p) = \big[-\lambda \big] 
\sum_q \big[-D(q) \big] \big[-G(p-q) \big] \big[-\Lambda(p-q,p) \big]. 
\nonumber 
\end{equation}
} 
is given as\footnote{
We use the thermal Green functions 
\begin{equation}
G({\bf p}, \tau) = - 
\big\langle T_\tau \big\{ c_{{\bf p}\sigma}(\tau) 
                          c_{{\bf p}\sigma}^\dag \big\} \big\rangle, 
\nonumber 
\end{equation}
for electrons and 
\begin{equation}
D({\bf q}, \tau) = - 
\big\langle T_\tau \big\{ \chi_{\bf q}(\tau) 
                          \chi_{\bf -q} \big\} \big\rangle, 
\nonumber 
\end{equation}
for bosons. 
The zeroth component of the four-momentum is given by 
$p_0 = -i \varepsilon_n$ or $q_0 = -i \omega_m$ as in [I]. 
The summation $\sum_q$ means $T\sum_m \sum_{\bf q}$. } 
\begin{equation}
\Sigma(p) = - \lambda \sum_q D(q) G(p-q) \Lambda(p-q,p). 
\label{int-self-energy} 
\end{equation}

\vskip 12mm

The Ward identity for electron-boson system 
is thoroughly clarified in the study of QED. 
\vskip 12mm
\begin{figure}[htbp]
\begin{center}
\includegraphics[width=8.0cm]{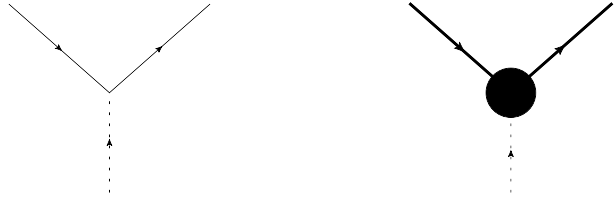}
\vskip 4mm
\caption{Free current vertex $\gamma^e$ (left) 
and full current vertex $\Gamma^e$ (right). 
The four-momentum of incoming/outgoing electron is $p$/$p+k$. }
\label{fig:CV}
\end{center}
\end{figure}
The four-divergence of the free vertex satisfies\footnote{
The free vertex is given as $\displaystyle \gamma_0^e(p+k,p)=e$ and 
$\displaystyle \gamma_\mu^e(p+k,p)
={e \over m}\big( p_\mu + { k_\mu \over 2} \big)$ 
for $\mu=1,2,3$ 
so that 
\begin{equation}
\sum_{\mu=0}^3 k_\mu \gamma_\mu^e(p+k,p) = 
e \big[ k_0 
     + {1 \over m}{\bf k}\cdot\big( {\bf p} + {{\bf k} \over 2} \big) \big]. 
\nonumber 
\end{equation}
One the other hand, 
\begin{equation}
G_0(p)^{-1}- G_0(p+k)^{-1} = k_0 
+ {1 \over m}{\bf k}\cdot\big( {\bf p} + {{\bf k} \over 2} \big), 
\nonumber 
\end{equation}
since 
\begin{equation}
G_0(p)^{-1} = - p_0 - \xi_{\bf p}. 
\nonumber 
\end{equation}
} 
\begin{equation}
\sum_{\mu=0}^3 k_\mu \gamma_\mu^e(p+k,p) = 
e G_0(p)^{-1}- e G_0(p+k)^{-1}. 
\label{Ward-QED-free} 
\end{equation}
The four-divergence of the full vertex satisfies 
the same relation as the free vertex 
\begin{equation}
\sum_{\mu=0}^3 k_\mu \Gamma_\mu^e(p+k,p) = 
e G(p)^{-1}- e G(p+k)^{-1}. 
\label{Ward-QED} 
\end{equation}
This relation for the full vertex is the Ward identity. 
If we decompose the full vertex as 
\begin{equation}
\Gamma_\mu^e(p+k,p) = \gamma_\mu^e(p+k,p) + {\tilde \Gamma}_\mu^e(p+k,p), 
\end{equation}
the interaction part satisfies 
\begin{equation}
\sum_{\mu=0}^3 k_\mu {\tilde \Gamma}_\mu^e(p+k,p) = 
e \Sigma(p+k)- e \Sigma(p). 
\label{Gamma-int} 
\end{equation}
Almost all of the contributions to the four-divergence 
of the full vertex cancel out 
so that only end-diagrams in Fig.~\ref{fig:QED} contain 
non-canceling contributions. 
The cancelation results from the four-divergence of the free vertex 
multiplied by $G_0(p)G_0(p+k)$ 
\begin{equation}
G_0(p) \big[ \sum_{\mu=0}^3 k_\mu \gamma_\mu^e(p+k,p) \big] G_0(p+k) 
= e G_0(p+k) - e G_0(p), 
\label{subtraction} 
\end{equation}
where the left-hand side is depicted 
as the left diagram in Fig.~\ref{fig:left}. 
\vskip 12mm
\begin{figure}[htbp]
\begin{center}
\includegraphics[width=9.0cm]{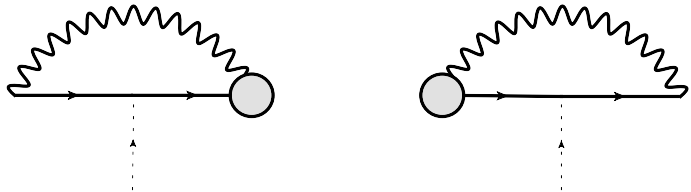}
\vskip 4mm
\caption{End-diagrams 
which contain the surviving contributions against the cancellation. 
The left end of each diagram links to the external electron propagator $G(p)$ 
and the right end to $G(p+k)$. }
\label{fig:QED}
\end{center}
\end{figure}
Only the contribution whose end electron propagator, 
which directly links to the external electron propagator 
(not shown in the figures) couples to the external field, 
has no canceling partner.\footnote{
See, for example, Peskin and Schroeder: 
{\it An Introduction to Quantum Field Theory} (Westview, Boulder, 1995). 
All the internal electron propagators in $\Gamma^e$ 
are classified into two types. 
One is the line and the other is the loop. 
Elementary canceling pair for the line is shown in Fig.~\ref{fig:cancel}. 
Such a pair for the loop is shown in Fig.~\ref{fig:coupling}. 
While the end-contributions remain from the line process, 
the similar contributions from the loop process cancel out 
after integrating over the internal variable of the loop. 
Thus we consider only end-contributions from the line process. } 
The left end-propagator which couples to the external field 
shown in Fig.~\ref{fig:left} 
is replaced by $ e G(p+k-q) $ 
shown in Fig.~\ref{fig:Dyson} 
after taking the four-divergence. 
\vskip 12mm
\begin{figure}[htbp]
\begin{center}
\includegraphics[width=10.0cm]{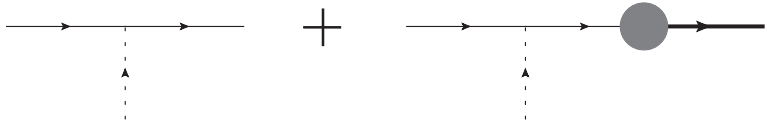}
\vskip 4mm
\caption{Left end-propagator. 
It is a part of the propagator coupling to external field 
used in the left diagram in Fig.~\ref{fig:QED}. 
Only the end free-propagator $G_0(p-q)$, 
which directly links to the external propagator $G_0(p)$, 
couples to external field in the left end-propagator. 
The gray circle is the electron self-energy $\Sigma$. 
Taking the four-divergence and using (\ref{subtraction}) 
we obtain two types of contributions. 
One which has $-G_0(p-q)$ at its left end 
is canceled if it is used in Fig.~\ref{fig:QED}. 
An example of the cancelation is shown in Fig.~\ref{fig:cancel}. 
The other which has $G_0(p+k-q)$ at its left end 
remains, because it has no canceling partner. }
\label{fig:left}
\end{center}
\end{figure}
\vskip 12mm
\begin{figure}[htbp]
\begin{center}
\includegraphics[width=7.0cm]{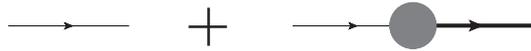}
\vskip 4mm
\caption{Remaining contribution of 
the left end-propagator in Fig.~\ref{fig:left} 
after taking the four-divergence. 
Every propagator carries the four-momentum $p+k-q$. 
Via Dyson equation it is identical to $G(p+k-q)$. }
\label{fig:Dyson}
\end{center}
\end{figure}
\vskip 12mm
\begin{figure}[htbp]
\begin{center}
\includegraphics[width=11.0cm]{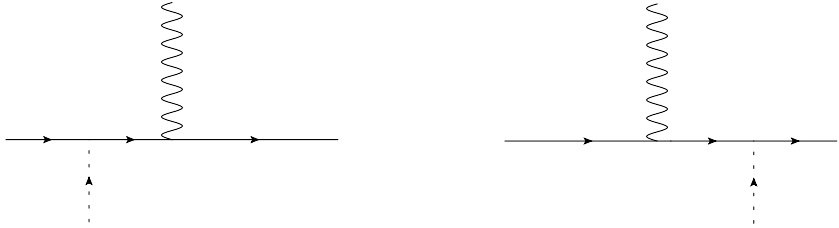}
\vskip 4mm
\caption{Elementary cancelation process between the canceling pair. 
The incoming electron, outgoing electron and boson 
carry $p$, $p+k-q$ and $q$ respectively. 
After taking the four-divergence 
the contribution of the left diagram reduces to 
$ -\lambda e \big[ G_0(p+k) - G_0(p) \big] D_0(q) G_0(p+k-q) $
and the right to 
$ -\lambda e G_0(p) D_0(q) \big[ G_0(p+k-q) - G_0(p-q) \big] $ 
by (\ref{subtraction}). 
The terms proportional to $G_0(p)D_0(q)G_0(p+k-q)$ cancel out each other. 
The other terms remain as the end-contribution. }
\label{fig:cancel}
\end{center}
\end{figure}
Thus the non-canceling contribution of the four-divergence 
of the left end-diagram in Fig.~\ref{fig:QED} becomes $ e \Sigma(p+k)$. 
By the same way 
the right end-diagram in Fig.~\ref{fig:QED} gives $ - e \Sigma(p)$. 
Consequently we obtain (\ref{Gamma-int}). 
\vskip 12mm
\begin{figure}[htbp]
\begin{center}
\includegraphics[width=11.0cm]{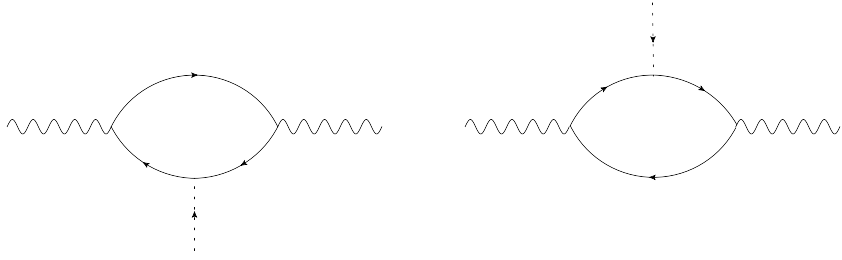}
\vskip 4mm
\caption{Pair of elementary loop cancelation. } 
\label{fig:coupling}
\end{center}
\end{figure}

\section{NCA for Charge Vertex}

Since the interaction vertex $\Lambda$ in Fig.~\ref{fig:Sigma} 
plays no crucial role to establish (\ref{Ward-QED}) 
as seen in the previous section, 
the Ward identity holds within the non-crossing approximation (NCA)\footnote{
The discussion in this section is based on \cite{Sch}. } 
which neglects the vertex correction to the electron-boson coupling 
($ \Lambda \rightarrow \lambda$). 

\vskip 12mm
\begin{figure}[htbp]
\begin{center}
\includegraphics[width=12.0cm]{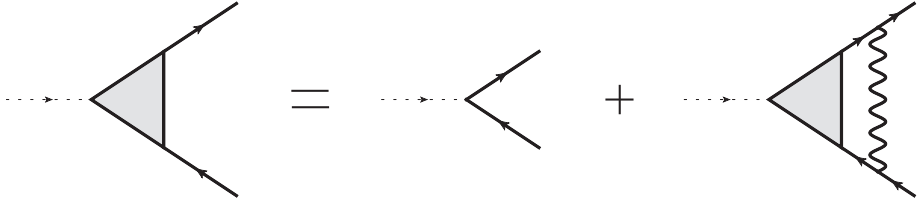}
\vskip 4mm
\caption{Ladder series for charge current vertex in NCA. }
\label{fig:NCA}
\end{center}
\end{figure}
The integral equation for the charge current vertex in NCA 
is depicted as Fig~\ref{fig:NCA} and given as\footnote{
The summation $\sum_{p'}$ means $T\sum_{n'} \sum_{{\bf p}'}$ 
as the footnote for (\ref{int-self-energy}). } 
\begin{equation}
\Gamma_\mu^e(p+k,p) = \gamma_\mu^e(p+k,p) - \lambda^2 \sum_{p'} 
G(p'+k) G(p') D(p-p') \Gamma_\mu^e(p'+k,p'). 
\label{int-Gamma} 
\end{equation}
The iterative solution of (\ref{int-Gamma}) reduces to the ladder series. 

To see the consistency between the Ward identity (\ref{Ward-QED}) 
and the ladder approximation (\ref{int-Gamma}), we use 
\begin{equation}
\sum_{\mu=0}^3 k_\mu \Gamma_\mu^e(p'+k,p') = 
e G(p')^{-1}- e G(p'+k)^{-1}, 
\end{equation}
in the four-divergence of (\ref{int-Gamma}) and obtain 
\begin{equation}
\sum_{\mu=0}^3 k_\mu \Gamma_\mu^e(p+k,p) = 
\sum_{\mu=0}^3 k_\mu \gamma_\mu^e(p+k,p) - \lambda^2 
e \sum_{p'} G(p'+k) D(p-p') + \lambda^2 e \sum_{p'} G(p') D(p-p'). 
\label{sum-Gamma'} 
\end{equation}
Using (\ref{Ward-QED-free}) and the self-energy in NCA 
\begin{equation}
\Sigma(p) = -\lambda^2 \sum_{p'} G(p') D(p-p'), 
\label{Sigma-NCA} 
\end{equation}
(\ref{sum-Gamma'}) is written as 
\begin{equation}
\sum_{\mu=0}^3 k_\mu \Gamma_\mu^e(p+k,p) = 
e G_0(p)^{-1}- e G_0(p+k)^{-1} + e \Sigma(p+k) - e \Sigma(p). 
\end{equation}
Using the Dyson equation 
\begin{equation}
G(p)^{-1} = G_0(p)^{-1}- \Sigma(p), 
\end{equation}
it is proven 
that the ladder approximation for the charge current vertex (\ref{int-Gamma}) 
satisfies the Ward identity (\ref{Ward-QED}). 
Thus the skeleton behavior of the Ward identity is seen in NCA. 

\section{NCA for Heat Vertex}

Since we can learn the skeleton behavior of the Ward identity 
from NCA as seen in the previous section, 
we investigate the integral equation for the heat current vertex 
within NCA.\footnote{
The discussion in this section is based on \cite{Ono}. }  

\begin{figure}[htbp]
\begin{center}
\includegraphics[width=8.0cm]{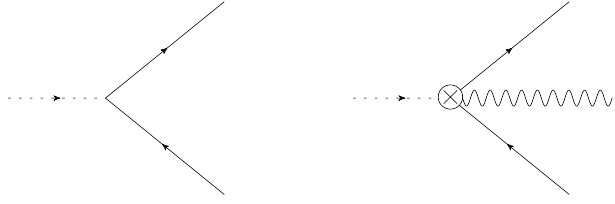}
\vskip 4mm
\caption{Diagrams for un-renormalized current vertex 
$ \gamma_\mu^Q $ (left) and $ \alpha_\mu^Q $ (right) 
in the case of electron-boson interaction. 
In the left diagram the incoming electron propagator is $G(p)$ and 
the outgoing one is $G(p+k)$. 
In the right diagram the pair of incoming and outgoing electron propagators is 
$G(p)G(p+k-q)$ or $G(p-q)G(p+k)$. } 
\label{fig:wave}
\end{center}
\end{figure}

Un-renormalized current vertexes depicted in Fig.~\ref{fig:J} 
are replaced by Fig.~\ref{fig:wave} in the case of electron-boson interaction. 
\vskip 12mm
\noindent
Since\footnote{
The Fourier transform of (88) in [I] leads to 
\begin{equation}
{\bf j}^Q({\bf -k}) = 
\lim_{\tau' \rightarrow \tau} {1 \over 2} 
\bigg( {\partial \over \partial \tau} - {\partial \over \partial \tau'} \bigg) 
{1 \over m}\sum_{\bf p} \Big( {\bf p} + { {\bf k} \over 2 } \Big) 
c_{{\bf p+k}\sigma}^\dag(\tau)  c_{{\bf p}\sigma}(\tau'). 
\nonumber
\end{equation}
In the case of $ K = K_0 + V_{\rm boson} $ 
the time-derivatives are calculated as 
\begin{equation}
{\partial \over \partial \tau} c_{{\bf p+k}\sigma}^\dag(\tau) 
= \big[ K, c_{{\bf p+k}\sigma}^\dag \big] 
= \xi_{\bf p+k} c_{{\bf p+k}\sigma}^\dag 
+ \lambda \sum_{\bf q} \chi_{\bf q} c_{{\bf p+k+q}\sigma}^\dag, 
\nonumber
\end{equation}
\begin{equation}
{\partial \over \partial \tau'} c_{{\bf p}\sigma}(\tau') 
= \big[ K, c_{{\bf p}\sigma} \big] 
= - \xi_{\bf p} c_{{\bf p}\sigma} 
- \lambda \sum_{\bf q} \chi_{\bf q} c_{{\bf p-q}\sigma}. 
\nonumber
\end{equation}
 } 
\begin{align}
{\bf j}^Q({\bf -k}) 
= {1 \over m} \sum_{\bf p} \sum_\sigma 
\Big( {\bf p} + { {\bf k} \over 2 } \Big) 
\Big[ & { \xi_{\bf p} + \xi_{\bf p+k} \over 2 } 
c_{{\bf p+k}\sigma}^\dag  c_{{\bf p}\sigma} 
\nonumber \\ 
+ & {\lambda \over 2} \sum_{\bf q} 
\big( \chi_{\bf -q} c_{{\bf p+k-q}\sigma}^\dag  c_{{\bf p}\sigma} 
    + \chi_{\bf q} c_{{\bf p+k}\sigma}^\dag  c_{{\bf p-q}\sigma} \big)
\Big], 
\label{j^Q(-k)} 
\end{align}
the vertexes are given as 
\begin{equation}
\gamma_\mu^Q(p+k,p) 
= {1 \over m}\Big( p_\mu + { k_\mu \over 2 } \Big) 
  { \xi_{\bf p} + \xi_{\bf p+k} \over 2 }, 
\ \ \ \ \ \ \ \ \ \ \ \ 
\alpha_\mu^Q 
= {\lambda \over 2m}\Big( p_\mu + { k_\mu \over 2 } \Big), 
\label{gamma-alpha} 
\end{equation}
for $\mu=1,2,3$. 
For the zeroth component in the limit of ${\bf k}\rightarrow 0$
\begin{equation}
\gamma_0^Q(p+k,p) 
= \xi_{\bf p}, 
\ \ \ \ \ \ \ \ \ \ \ \ 
\alpha_0^Q 
= \lambda, 
\label{gamma-alpha-0} 
\end{equation}
from (67) in [I]. 

\begin{figure}[htbp]
\begin{center}
\includegraphics[width=9.0cm]{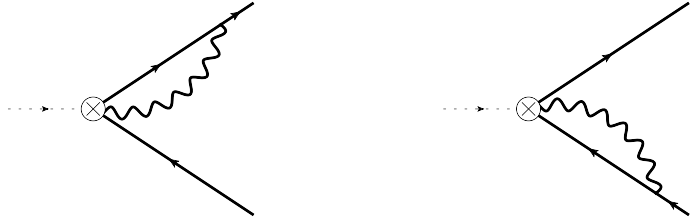}
\vskip 4mm
\caption{Non-crossing contributions to the heat current vertex 
besides the ladder series. 
The process at the left vertex is 
$\chi_{\bf -q} c_{{\bf p+k-q}\sigma}^\dag  c_{{\bf p}\sigma}$. 
The process at the right vertex is 
$\chi_{\bf q} c_{{\bf p+k}\sigma}^\dag  c_{{\bf p-q}\sigma}$. } 
\label{fig:SV}
\end{center}
\end{figure}

Even within NCA 
we have to consider the contributions, 
besides the ladder series in Fig.~\ref{fig:NCA}, 
depicted in Fig.~\ref{fig:SV} 
in the integral equation for the vertex. 
\vskip 12mm
\noindent
Thus the integral equation becomes 
\begin{align}
\Gamma_\mu^Q(p+k,p) = \gamma_\mu^Q(p+k,p) 
& - \lambda^2 \sum_{p'} G(p'+k) G(p') D(p-p') \Gamma_\mu^Q(p'+k,p') 
\nonumber \\ 
& - \alpha_\mu^Q \lambda \sum_{\bf q}
\big[ G(p+k-q)D(q) + G(p-q)D(q) \big]. 
\label{int-Q-alpha} 
\end{align}
Taking the four-divergence and using the Ward identity 
\begin{equation}
\sum_{\mu=0}^3 k_\mu \Gamma_\mu^Q(p'+k,p') = 
p_0' G(p'+k)^{-1} - (p_0'+k_0) G(p')^{-1}, 
\label{full-W-heat} 
\end{equation}
(\ref{gamma-alpha}) and (\ref{gamma-alpha-0}) 
we obtain 
\begin{align}
\sum_{\mu=0}^3 & k_\mu \Gamma_\mu^Q(p+k,p) = 
k_0 \xi_{\bf p} 
+ {1 \over 2m} {\bf k}\cdot\Big( {\bf p}+{ {\bf k}\over 2 }\Big) 
\big[  \xi_{\bf p} + \xi_{\bf p+k} + \Sigma(p) + \Sigma(p+k) \big] 
\nonumber \\ 
+ & k_0 \big[ \Sigma(p) + \Sigma(p+k) \big] 
- \lambda^2 \sum_{p'} D(p-p') 
\big[ p_0' G(p') - (p_0'+k_0)G(p'+k) \big], 
\label{div-Gamma} 
\end{align}
where the self-energy is given by (\ref{Sigma-NCA}). 

Using the Jonson-Mahan transmutation \cite{JM} 
\begin{equation}
\xi_{\bf p} + \Sigma(p) = -p_0 - G(p)^{-1}, 
\end{equation}
the first line of the right-hand side of (\ref{div-Gamma}) 
is written as 
\begin{equation}
k_0 \xi_{\bf p} 
- {1 \over 2m} {\bf k}\cdot\Big( {\bf p}+{ {\bf k}\over 2 }\Big) 
\big[ 2p_0 + k_0 + G(p)^{-1} + G(p+k)^{-1} \big], 
\label{line-1} 
\end{equation}
and $k_0$ in [ ] is negligible 
in the limit of $k_0 \rightarrow 0$. 
We expect that this line is equal to 
\begin{equation}
p_0 G_0(p+k)^{-1} - (p_0+k_0) G_0(p)^{-1} 
= k_0 \xi_{\bf p} 
- {1 \over m} {\bf k}\cdot\Big( {\bf p}+{ {\bf k}\over 2 }\Big) p_0. 
\label{Ward-Q0} 
\end{equation}
But there is no counter term to cancel out 
the term proportional to $ G(p)^{-1} + G(p+k)^{-1} $ in (\ref{line-1}) 
within NCA. 
Since it does not contribute to the conductivity, 
this unnecessary term was neglected in \cite{JM}. 
However, we should find the counter term to establish the Ward identity. 
This task is accomplished by the exact diagrammatics in the next section. 
In the following discussion within this section 
we neglect this unnecessary term. 

It should be noted that the relation 
\begin{equation}
\sum_{\mu=0}^3 k_\mu \gamma_\mu^Q(p+k,p) = 
  p_0 G_0(p+k)^{-1} - (p_0+k_0) G_0(p)^{-1}, 
\label{wrong} 
\end{equation}
that is the same relation 
as satisfied by the full quantities (\ref{full-W-heat}), 
does not hold\footnote{
This fact is not widely recognized. 
Even in \cite{Ono} (\ref{wrong}) is assumed 
so that the definition of $\gamma_\mu^Q$ (4$\cdot$13a) is incorrect. 
Since we consider the limit of ${\bf k}\rightarrow 0$, 
neglecting the difference between $\xi_{\bf p+k}$ and $\xi_{\bf p}$ 
(\ref{div-gammaQfree}) gives 
\begin{equation}
\sum_{\mu=0}^3 k_\mu \gamma_\mu^Q(p+k,p) = \xi_{\bf p} 
\Big[ k_0 
    + {1 \over m} {\bf k}\cdot\Big( {\bf p}+{ {\bf k}\over 2 }\Big) \Big] 
= \xi_{\bf p} \big[ G_0(p)^{-1} - G_0(p+k)^{-1} \big]. 
\nonumber 
\end{equation}
This relation is parallel to 
\begin{equation}
\sum_{\mu=0}^3 k_\mu \gamma_\mu^e(p+k,p) = 
e \big[ G_0(p)^{-1}- G_0(p+k)^{-1} \big]. 
\nonumber 
\end{equation}
} 
in contrast to (\ref{Ward-QED-free}). 
Actually 
\begin{equation}
\sum_{\mu=0}^3 k_\mu \gamma_\mu^Q(p+k,p) = 
k_0 \xi_{\bf p} 
+ {1 \over 2m} {\bf k}\cdot\Big( {\bf p}+{ {\bf k}\over 2 }\Big) 
( \xi_{\bf p} + \xi_{\bf p+k} ). 
\label{div-gammaQfree} 
\end{equation}
The right-hand side of (\ref{Ward-Q0}) results from (\ref{div-gammaQfree}) 
plus the contribution depicted in Fig.~\ref{fig:SV}. 

\begin{figure}[htbp]
\begin{center}
\includegraphics[width=11.0cm]{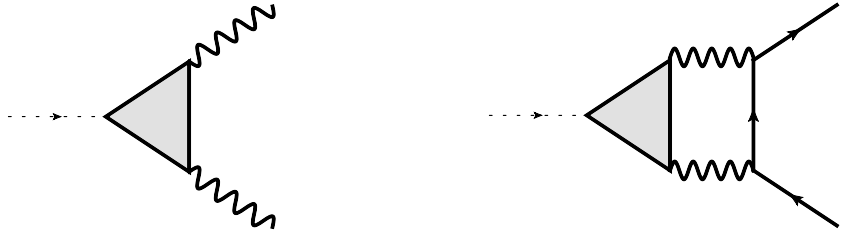}
\vskip 4mm
\caption{Heat current vertex for bosons (left) and 
vertex correction for electrons (right). 
Both processes vanish in the case of charge current vertex, 
since the boson is charge neutral and does not carry charge. } 
\label{fig:NCA-fluc}
\end{center}
\end{figure}

The last term in the right-hand side of (\ref{div-Gamma}) 
is written as 
\begin{equation}
p_0 \Sigma(p) - (p_0+k_0)\Sigma(p+k) 
+ \lambda^2 \sum_{q} \big[ q_0 D(q) - (q_0+k_0) D(q+k) \big] G(p-q), 
\label{Sigma+coupling} 
\end{equation}
where the self-energy is given by (\ref{Sigma-NCA}). 
Since we need only $ p_0 \Sigma(p) - (p_0+k_0)\Sigma(p+k) $ 
in (\ref{Sigma+coupling}), we should have included\footnote{
In \cite{Ono} a coupled equation for $\Gamma_\mu^Q$ and $\Delta_\mu^Q$ 
is discussed in the case of electron-phonon system. See Fig.~2 there. } 
\begin{equation}
- \lambda^2 \sum_{q} D(q) D(q+k) G(p-q) \Delta_\mu^Q(q+k,q), 
\label{wave-coupling} 
\end{equation}
in (\ref{int-Q-alpha}).\footnote{
Assuming that $\Delta_\mu^Q$ in NCA satisfies the Ward identity, 
(144) in [I], 
\begin{equation}
\sum_{\mu=0}^3 k_\mu \Delta_\mu^Q(q+k,q) = 
q_0 D(q+k)^{-1} - (q_0+k_0) D(q)^{-1}, 
\nonumber
\end{equation}
the four-divergence of (\ref{wave-coupling}) reduces to 
\begin{equation}
- \lambda^2 \sum_{q} \big[ q_0 D(q) - (q_0+k_0) D(q+k) \big] G(p-q), 
\end{equation}
and cancels out the last term in (\ref{Sigma+coupling}). } 
The vertex for bosons $\Delta_\mu^Q$ in NCA 
is depicted in Fig.~\ref{fig:NCA-fluc}-(left) 
and (\ref{wave-coupling}) is depicted in Fig.~\ref{fig:NCA-fluc}-(right). 
Even if the direct coupling of bosons to external field is absent, 
the intermediate electron-hole pair state in the boson propagator 
couples to external field\footnote{
In the case of charge current vertex 
the coupling of bosons to external field vanishes, 
because the boson is charge neutral and does not carry charge. 
On the other hand, bosons carry heat so that $\Delta^Q$ is non-zero. 
See arXiv:1109.1404 for detail. } 
as depicted in Fig.~\ref{fig:coupling} 
so that non-zero $\Delta_\mu^Q$ arises. 

\vskip 12mm

By putting above considerations together 
we manage to reach the expected form of the Ward identity 
\begin{align}
\sum_{\mu=0}^3 k_\mu \Gamma_\mu^Q(p+k,p) &= 
  p_0 G_0(p+k)^{-1} - (p_0+k_0) G_0(p)^{-1} 
- p_0 \Sigma(p+k)   + (p_0+k_0) \Sigma(p) 
\nonumber \\ 
&= p_0 G(p+k)^{-1} - (p_0+k_0) G(p)^{-1}. 
\end{align}
However, this discussion is incomplete, 
since I have shut my eyes to the unnecessary term.\footnote{
Although \cite{Ono} is written under the assumption 
that the Ward identity for the heat current holds within NCA, 
it is not the case. } 

\section{Diagrammatics for Heat Vertex}

Now the skeleton behavior of the heat current vertex 
has been clarified in NCA. 
However, the unnecessary term remains 
so that the Ward identity is not completely satisfied within NCA. 
Thus we shall show the complete explanation of the Ward identity 
by exact diagrammatics in this section. 

\begin{figure}[htbp]
\begin{center}
\includegraphics[width=11.0cm]{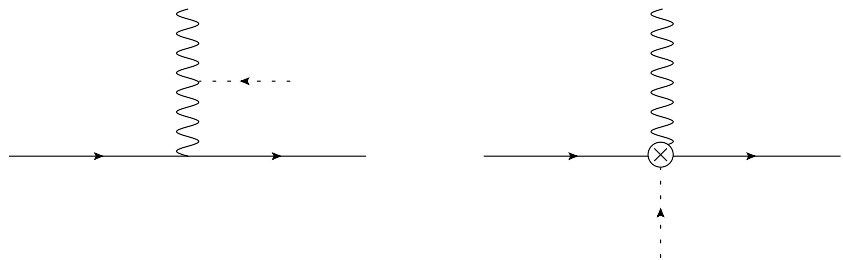}
\vskip 4mm
\caption{Processes necessary to form the canceling quartet. } 
\label{fig:cancel-wave}
\end{center}
\end{figure}

In the case of the line-diagram 
we should add the processes in Fig.~\ref{fig:cancel-wave} 
to the canceling pair in Fig.~\ref{fig:cancel} 
to form the canceling quartet. 
\vskip 12mm
\noindent
The renormalized vertex ${\tilde \gamma}_\mu^Q$ defined 
as the sum of the free vertex $\gamma_\mu^Q$ 
and the processes in Fig.~\ref{fig:SV} with $\mu=1,2,3$, satisfies 
\begin{equation}
\sum_{\mu=0}^3 k_\mu {\tilde \gamma}_\mu^Q(p+k,p) = 
  p_0 G_0(p+k)^{-1} - (p_0+k_0) G_0(p)^{-1}, 
\label{renormalized-gamma} 
\end{equation}
as the discussion in the previous section.\footnote{
The self-energy in the skeleton analysis of NCA is (\ref{Sigma-NCA}) 
but the fully renormalized self-energy used in this section 
is (\ref{int-self-energy}). } 
Since the {\bf disposal}\footnote{
By the renormalization procedure $\gamma^Q \rightarrow {\tilde \gamma}^Q$ 
an unnecessary term 
\begin{equation}
{1 \over 2m} {\bf k}\cdot\Big( {\bf p}+{ {\bf k}\over 2 }\Big) 
G(p) \lambda D(q) G(p+k-q), 
\end{equation}
arises from Fig.~\ref{fig:cancel}-(left) via 
\begin{equation}
-{1 \over 2m} {\bf k}\cdot\Big( {\bf p}+{ {\bf k}\over 2 }\Big) 
\big[ G(p)^{-1} + G(p+k)^{-1} \big] (-1)^5 
G(p) G(p+k) \lambda D(q) G(p+k-q). 
\end{equation}
It is canceled out by the three-divergence 
of Fig.~\ref{fig:cancel-wave}-(right) 
\begin{equation}
{\lambda \over 2m} {\bf k}\cdot\Big( {\bf p}+{ {\bf k}\over 2 }\Big) 
(-1)^3 G(p) D(q) G(p+k-q). 
\end{equation}
Here I have used the full propagators $G$ and $D$. } 
of the unnecessary term is possible 
in the exact diagrammatics, (\ref{renormalized-gamma}) holds exactly. 
Corresponding to (\ref{subtraction}) 
the four-divergence of the coupling to external field 
in Fig.~\ref{fig:cancel}-(left) satisfies 
\begin{equation}
G_0(p) \big[ \sum_{\mu=0}^3 k_\mu {\tilde \gamma}_\mu^Q(p+k,p) \big] G_0(p+k) 
= p_0 G_0(p) - (p_0+k_0) G_0(p+k), 
\end{equation}
when we use the renormalized vertex.
By the same manner 
the renormalized coupling in Fig.~\ref{fig:cancel-wave}-(left) satisfies 
\begin{equation}
D_0(q) \big[ \sum_{\mu=0}^3 k_\mu {\tilde \beta}_\mu^Q(q+k,q) \big] D_0(q+k) 
= q_0 D_0(q) - (q_0+k_0) D_0(q+k). 
\end{equation}

\begin{figure}[htbp]
\begin{center}
\includegraphics[width=3.0cm]{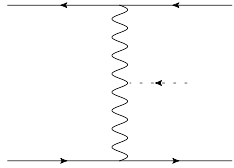}
\vskip 4mm
\caption{Coupling of boson to external field. } 
\label{fig:up-down}
\end{center}
\end{figure}

After taking the four-divergence 
the process in Fig.~\ref{fig:cancel}-(left) gives\footnote{
Before taking the four-divergence 
the process in Fig.~\ref{fig:cancel}-(left) gives 
\begin{equation}
\big[-\lambda \big] 
\big[-G_0(p) \big] {\tilde \gamma}_\mu^Q \big[-G_0(p+k) \big] 
\big[-D_0(q) \big] \big[-G(p+k-q) \big], 
\nonumber 
\end{equation}
and the process in Fig.~\ref{fig:cancel-wave}-(right) gives 
\begin{equation}
\lambda 
\big[-G_0(p) \big] \big[-D_0(q) \big] \big[-G_0(p+k-q) \big], 
\nonumber 
\end{equation}
which results from the coupling $\alpha_0^Q$. } 
\begin{equation}
- \lambda 
\big[ p_0 G_0(p) - (p_0+k_0) G_0(p+k) \big] D_0(q) G_0(p+k-q), 
\end{equation}
the process in Fig.~\ref{fig:cancel}-(right) gives 
\begin{equation}
- \lambda 
G_0(p) D_0(q) \big[ (p_0-q_0) G_0(p-q) - (p_0+k_0-q_0) G_0(p+k-q) \big], 
\end{equation}
the process in Fig.~\ref{fig:cancel-wave}-(left) gives\footnote{
See Fig.~\ref{fig:up-down} for a simple example. 
The term proportional to $q_0D(q)$ 
forms the canceling quartet concerning the lower electron line. 
On the other hand, the term proportional to $(q_0-k_0)D(q-k)$ 
forms the canceling quartet concerning the upper electron line. } 
\begin{equation}
- \lambda
G_0(p) \big[ (q_0-k_0) D_0(q-k) - q_0 D_0(q) \big] G_0(p+k-q), 
\end{equation}
and the process in Fig.~\ref{fig:cancel-wave}-(right) gives\footnote{
As discussed in the footnote for the {\bf disposal}, 
the three-divergence is renormalized into ${\tilde \gamma}^Q$ 
so that the remaining is the zeroth component only. } 
\begin{equation}
- \lambda k_0 
G_0(p) D_0(q) G_0(p+k-q). 
\end{equation}
Then the terms proportional to $G_0(p)D_0(q)G_0(p+k-q)$ 
in these four equations cancel out. 
\vskip 12mm

In the case of the loop-diagram 
the cancelation is not complete\footnote{
For example, 
when we discuss the cancelation between two processes 
in Fig.~\ref{fig:coupling}, 
both left and right bosons, which couple to electron loop, 
can form the canceling quartet 
and the excess contributions survive against the cancelation. 
See also the footnote for non-zero $\Delta^Q$. } 
so that the external field couples to bosons 
via indirect processes as shown in Fig.~\ref{fig:coupling}. 

\begin{figure}[htbp]
\begin{center}
\includegraphics[width=9.0cm]{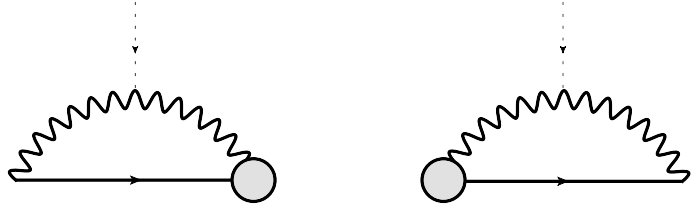}
\vskip 4mm
\caption{End-diagrams 
in terms of boson propagator. }
\label{fig:fluc}
\end{center}
\end{figure}

\begin{figure}[htbp]
\begin{center}
\includegraphics[width=9.0cm]{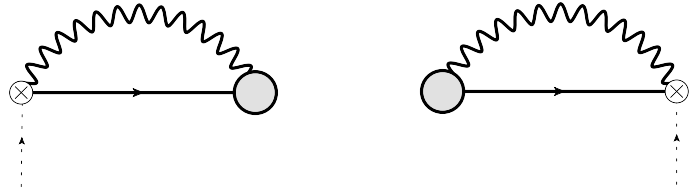}
\vskip 4mm
\caption{End-diagrams 
in terms of $\alpha_\mu^Q$. }
\label{fig:heat-Sigma}
\end{center}
\end{figure}

Since the elementary cancelation mechanism has become clear 
by the above explanation, 
we can discuss the Ward identity 
by the same procedure as in the case of the charge current vertex. 
Namely, we only have to consider the end-contributions\footnote{
The contribution via the violation of the loop cancelation 
is taken into account by the coupling of bosons to external field. 
Such a coupling is absent in the case of charge current 
because of the loop cancelation. } 
to establish the relation corresponding to (\ref{Gamma-int}). 
The end-contribution from the left diagram in Fig.~\ref{fig:QED} is 
\begin{equation}
- (p_0+k_0) \Sigma(p+k) + \lambda \sum_q q_0 D(q) G(p+k-q) \Lambda(p+k-q,p+k), 
\label{End-1} 
\end{equation}
and the one from the right diagram is 
\begin{equation}
p_0 \Sigma(p) - \lambda \sum_q q_0 D(q) G(p-q) \Lambda(p,p-q). 
\label{End-2} 
\end{equation}
The terms proportional to $q_0$ in these contributions 
are canceled out\footnote{
To see the cancelation 
the second term in (\ref{End-1}) is written as 
\begin{equation}
\lambda \sum_{q'} (q_0'+k_0) D(q'+k) G(p-q') \Lambda(p-q',p+k), 
\nonumber 
\end{equation}
by the variable change $q-k \rightarrow q'$. } 
by the end-contributions in terms of boson propagator; 
\begin{equation}
- \lambda \sum_q (q_0+k_0) D(q+k) G(p-q) \Lambda(p-q,p+k), 
\label{End-3} 
\end{equation}
from the process in Fig.~\ref{fig:fluc}-(left) 
and 
\begin{equation}
\lambda \sum_q q_0 D(q) G(p-q) \Lambda(p,p-q), 
\label{End-4} 
\end{equation}
from the process in Fig.~\ref{fig:fluc}-(right). 
\vskip 12mm
\noindent
Thus the sum of (\ref{End-1}), (\ref{End-2}), (\ref{End-3}) and (\ref{End-4}) 
results in 
\begin{equation}
p_0 \Sigma(p) - (p_0+k_0) \Sigma(p+k). 
\label{End-Sigma} 
\end{equation}
The end-contributions in terms of $\alpha_\mu^Q$ 
depicted in Fig.~\ref{fig:heat-Sigma} are\footnote{
Since the three-divergence is already renormalized into ${\tilde \gamma}^Q$, 
we only consider the zeroth component. } 
\begin{equation}
k_0 \big[ \Sigma(p) + \Sigma(p+k) \big]. 
\label{End-alpha} 
\end{equation}
\vskip 12mm
\noindent
From (\ref{End-Sigma}) plus (\ref{End-alpha}) we obtain 
\begin{equation}
(p_0+k_0)\Sigma(p) - p_0\Sigma(p+k). 
\label{End+End} 
\end{equation}

From (\ref{renormalized-gamma}) and (\ref{End+End}) 
we obtain the Ward identity 
\begin{equation}
\sum_{\mu=0}^3 k_\mu \Gamma_\mu^Q(p+k,p) = 
   p_0      \big[ G_0(p+k)^{-1} - \Sigma(p+k) \big] 
- (p_0+k_0) \big[ G_0(p)^{-1}   - \Sigma(p)   \big], 
\end{equation}
for heat current vertex 
It should be noted that 
the interacting process in Fig.~\ref{fig:SV} plays a dual role. 
One is to establish (\ref{renormalized-gamma}) and the other (\ref{End+End}). 
Thus (\ref{renormalized-gamma}) does not hold for free electrons.  

\section{Remarks}

The sections of {\bf Exercise} and {\bf Acknowledgements} 
are common to [I] so that I do not repeat here. 
Some typographic errors in [I] have been listed 
in the section of {\bf Remarks} in [II]. 
The errors in [II] should be corrected as follows. 

\begin{list}{}{}
\item[$\bullet$] The first line of \S 4 should be 
\lq\lq The linearized GL transport theory in 2D $\cdot\cdot\cdot$ ". 
\item[$\bullet$] The first line of \S 5 should be 
\lq\lq The calculations for electrons in \S 3 are translated into those [2] 
for Cooper pairs$^{30}$ straightforwardly.$^{31}$"
\end{list}

Exploiting this opportunity 
I add the following explanation to (147) and (149) in [I]. 
When we put $k_0=0$ 
\begin{equation}
\sum_{\mu=1}^3 k_\mu \Gamma_\mu^e \fallingdotseq e \Big\{ 
\big[ \xi_{\bf p+k} + \Sigma'(p+k) \big] - 
\big[ \xi_{\bf p} + \Sigma'(p) \big] \Big\} 
\fallingdotseq {e \over m^*}{\bf k}\cdot{\bf p}, 
\nonumber 
\end{equation}
where $\Sigma'(p)$ is the real part of $\Sigma(p)$ and 
$\displaystyle \xi_{\bf p} + \Sigma'(p) \fallingdotseq 
{{\bf p}^2 \over 2m^*} - \mu^*$. Thus we obtain (147) in [I]. 
Similarly 
\begin{equation}
\sum_{\mu=1}^3 k_\mu \Gamma_\mu^Q \fallingdotseq - p_0 \Big\{ 
\big[ \xi_{\bf p+k} + \Sigma'(p+k) \big] - 
\big[ \xi_{\bf p} + \Sigma'(p) \big] \Big\} 
\fallingdotseq -{p_0 \over m^*}{\bf k}\cdot{\bf p}, 
\nonumber 
\end{equation}
leads to (149) in [I] with $k_0=0$. 


\end{document}